\documentclass[preprint2]{aastex}

\usepackage{natbib}
\usepackage[colorlinks,
            linkcolor=red,
            anchorcolor=blue,
            citecolor=blue
            ]{hyperref}

\shorttitle{Chirality and Magnetic Configurations of Solar Filaments}
\shortauthors{Ouyang, Zhou, Chen, \& Fang}

\begin{document}
%\nocite{*}
\bibliographystyle{yahapj}

\title{Chirality and Magnetic Configurations of Solar Filaments}

\author{Y. Ouyang\altaffilmark{1,2}, Y. H. Zhou\altaffilmark{1,3},
P. F. Chen\altaffilmark{1,3}, and C. Fang\altaffilmark{1,3}}

\affil{$^1$ School of Astronomy \& Space Science, Nanjing University,
        Nanjing 210023, China; \href{mailto:chenpf@nju.edu.cn}{chenpf@nju.edu.cn}}
     %   Nanjing 210023, China; \email{chenpf@nju.edu.cn}}
\affil{$^2$ School of Science, Linyi University, Linyi 276000, China}
\affil{$^3$ Key Lab of Modern Astron. \& Astrophys. (Ministry of
        Education), Nanjing University, China}

\begin{abstract}
It has been revealed that the magnetic topology in the solar atmosphere
displays hemispheric preference, i.e., negative/positive helicity in the
northern/southern hemisphere, respectively. However, the strength of the
hemispheric rule and its cyclic variation are controversial. In this paper,
we apply a new method based on filament drainage to 571 erupting filaments
from 2010 May to 2015 December in order to determine the filament chirality and
its hemispheric preference.  It is found that 91.6\% of our sample of erupting
filaments follow the hemispheric rule of helicity sign. It is found that the
strength of the hemispheric preference of the quiescent filaments decreases
slightly from $\sim$97\% in the rising phase to $\sim$85\% in the declining
phase of solar cycle 24, whereas the strength of the intermediate filaments
keeps a high value around $96\pm4\%$ all the time. Only the active-region
filaments show significant variations. Their strength of the hemispheric rule
rises from $\sim$63\% to $\sim$95\% in the rising phase, and keeps a high value
$82\pm5\%$ during the declining phase. Furthermore, during a half-year period
around the solar maximum, their hemispheric preference totally vanishes.
Besides, we also diagnose the magnetic configurations of the filaments based on
our indirect method, and found that in our sample of erupting events, 89\% are
inverse-polarity filaments with a flux rope magnetic configuration, whereas
11\% are normal-polarity filaments with a sheared arcade configuration.
\end{abstract}

\keywords{Sun: activity --- Sun: magnetic fields --- Sun: filaments, prominences}

\section{Introduction}

The magnetic field in the solar atmosphere, which is responsible for solar
eruptions \citep{Che11}, originates from the tachocline at the bottom of the
convection zone, and rises up buoyantly out of the solar surface. Because of
the antisymmetry of the Coriolis force, the magnetic structures in the solar
atmosphere often display symmetry or antisymmetry between the northern and the
southern hemispheres, i.e., the $\alpha$-effect. A typical example is the
Joy's law \citep{dsil93, wang91, fan94}. On the other hand, the interaction
between the buoyant flux tubes and the turbulent convection zone may weaken
any hemispheric preference, i.e., the $\Sigma$-effect \citep{long98}. In
addition, there are other effects such as the surface motions and magnetic
diffusion. As a result, the strength of the hemispheric preference is expected
to vary for different proxies of the magnetic topology. For example, in terms
of the current helicity, it was found that about $70\pm12\%$ of active regions
follow the hemispheric rule \citep{See90, Pev95, Abr97, Bao98, Hag05, Zha06,
Hao11, Gos13, Liu14}, i.e., the helicity is negative in the northern hemisphere
and positive in the southern hemisphere. In terms of the sunspot whorls,
although as high as 80\% of the H$\alpha$ whorls were found to be
counterclockwise (measured inwardly) in the northern hemisphere and clockwise
in the southern hemisphere \citep{Hal27, Ric41, Din87}. However, it was noticed
that only a small fraction of sunspots show a vortical structure \citep{Ric41},
and most sunspot whorls include both clockwise and counterclockwise H$\alpha$
fibrils \citep{Pev03a}. A similar complex situation happens to the X-ray
sigmoids \citep{Rus96, Lim09}, since they suffer from projection effects and
even potential magnetic fields may also have similar shapes.

Another important proxy of the magnetic topology is the chirality of filament
channels and their overlying coronal arcade, as proposed by \citet{Mar92}. Here 
the chirality of a filament channel is defined to be dextral/sinistral if the
axial magnetic field of the filament is to the right/left when viewed from the
positive polarity side of the filament channel. The chirality of 
filament channels originates from several sources, e.g., the twist of the 
magnetic field before emerging into the solar atmosphere, the interaction
between two neighboring flux systems, and solar surface motions \citep{vanb89,
zirk97, vanb98, mack00}. These sources are further divided into 8 mechanisms,
each of which produces the chirality pattern consistent and/or inconsistent
with the hemispheric rule \citep{yeat09}. Therefore, it is important to
determine the percentage of solar filaments which follow the hemispheric
rule. With a sample of 73 quiescent filaments, they found that $\sim$80\% of 
them were either dextral (with negative helicity) in the northern hemisphere 
or sinistral (with positive helicity) in the southern hemisphere. In their 
paper, they also proposed a rule, called Martin's rule hereafter, i.e., a 
dextral filament has right-bearing barbs, whereas a sinistral filament has 
left-bearing barbs. With such a one-to-one correspondence, one can immediately 
determine the chirality of a filament by looking at the H$\alpha$ image without 
the help of vector magnetograms. Applying this rule to the H$\alpha$ images 
during 2000--2001, \citet{Pev03b} confirmed that $\sim$83$\pm3\%$ of the 
quiescent filament follow the hemispheric rule of helicity. Compared to 
quiescent filaments, active-region filaments seem to have a weaker hemispheric 
preference in helicity. For instance, \citet{Pev03b} found that only 
$\sim$76$\pm1\%$ of active-region filaments follow the hemispheric rule. As an 
extreme result, \citet{Mar94} stated that active-region filaments do not obey 
the hemispheric rule, with nearly half filaments being dextral or sinistral in 
each hemisphere. One might argue that the contradictory results between
\citet{Mar94} and \citet{Pev03b} are due to the small sample, e.g., only 31
filaments in the former. However, later investigations with bigger samples
also led to opposing conclusions. For example, with 123 filaments,
\citet{Yea07} found that 82\% of all filaments follow the hemispheric rule,
whereas \citet{Ber05} identified the chirality of 658 filaments with an
automated detection method, and found that only 68\% of them obey the
hemispheric rule. Such a result of weak preference might be due to the 
limited resolution of the H$\alpha$ full-disk observations so that the 
automated method cannot work well. What is even worse, \citet{mart14} claimed 
that the hemispheric preference seems to disappear or reverse during parts of 
the declining phase.

The discrepancy between different researchers might be partly 
attributed to their chirality identification method, where the filament 
chirality is determined by the bearing sense of the filament barbs according to 
Martin's rule. The resulting statistics is contaminated by three factors: (1) 
Projection effects: A left-bearing barb might be falsely identified to be 
right-bearing due to the projection effects. This problem becomes more serious 
as the filament is closer to the solar limb; (2) Limitation of Martin's rule: 
As pointed out by \citet{guo10} and \citet{Che14}, Martin's rule is applicable 
to the filaments supported by a magnetic flux rope only. For filaments which 
are supported by a magnetic sheared arcade, the correspondence between the 
filament chirality and the bearing sense of the filament barbs would be 
opposite to the Martin's rule; (3) Multi-sampling: The typical lifetime of a 
filament is weeks or even months. Therefore, some filaments might be counted 
several times depending on their lifetime. In order to determine the strength 
of the hemispheric preference of the filament chirality, a better chirality 
identification method should be applied. For example, \citet{shee13} 
proposed to use the ``coronal cells" in the extreme ultra-violet (EUV) images 
of the Sun along with photospheric magnetograms to determine the chirality of 
the filament channel. This method is applicable when the coronal cells are 
clear.

Recently, \citet{Che14} proposed an indirect method to determine the chirality
of a filament without the help of vector magnetograms, which is independent of
the magnetic type of the filament, i.e., of normal-polarity or 
inverse-polarity. This method is based
on the observational fact that when a filament erupts, parts of the cold
plasmas drain down along the two legs of the supporting magnetic field lines
and impact the solar surface, forming two conjugate draining sites
\citep{Zho06, Tri13, Che14}. With respect to the magnetic polarity
inversion line (PIL) or the filament spine, the conjugate draining sites are
either left-skewed or right-skewed, corresponding to dextral or sinistral
chirality, respectively. One of the advantages of this method is that the
draining sites are well separated and are close to the solar surface (i.e.,
not suspended in the corona as the filament barbs), which do not suffer from
the projection effects. Moreover, only the erupting stage of a filament is
considered, by which a filament is sampled only once.

In this paper, we attempt to apply this new method to examine the strength of
the hemispheric preference of filament chirality. The paper is organized as
follows: The data sample and the chirality identification method are described
in \S\ref{sec2}, the results are presented in \S\ref{sec3}, which are discussed
in \S\ref{sec4} before a summary in \S\ref{sec5}.

\section{Data Sampling and Analysis}\label{sec2}

The {\em Solar Dynamics Observatory} ({\it SDO}) mission provides
high-resolution EUV images and magnetograms, which are observed by the 
Atmospheric Imaging Assembly \citep[AIA,][]{Lem12} and the Helioseismic and 
Magnetic Imager \citep[HMI,][]{Sch12}, respectively. Since its launch in early 
2010, the satellite has been monitoring the Sun continuously, covering both the 
rising and the beginning of the declining phases of solar cycle 24. 
Solar activities, including filament eruptions, were routinely recorded by
the Heliophysics Event Knowledgebase \citep[HEK,][]{hurl12}. From 2010 May 13
to 2015 December 31, there are more than 1000 erupting filaments/prominences.
Roughly half of these events are prominences above or behind the limb
\citep{mcca15}. With all these events excluded, 576 filaments are found to
erupt on the disk. Among these events, only 5 erupting filaments have no clear
draining sites. Therefore, a total of 571 erupting filaments are selected as
our sample. The {\it SDO}/AIA observes the Sun in seven EUV and three UV
channels with a pixel size of $0\farcs 6$ and a high time cadence of 12 s. In 
this study we use the 304 \AA, 171 \AA, and 193 \AA\ bandpasses in order to 
trace the filament eruptions and the brightenings associated with the filament 
drainage. These filament eruptions are also monitored by the {\it Global 
Oscillation Network Group} ({\it GONG}) in H$\alpha$ \citep{Har11}, where 
filament barbs can be clearly recognized in many of them.

\begin{figure*}
\epsscale{2.1}
\plotone{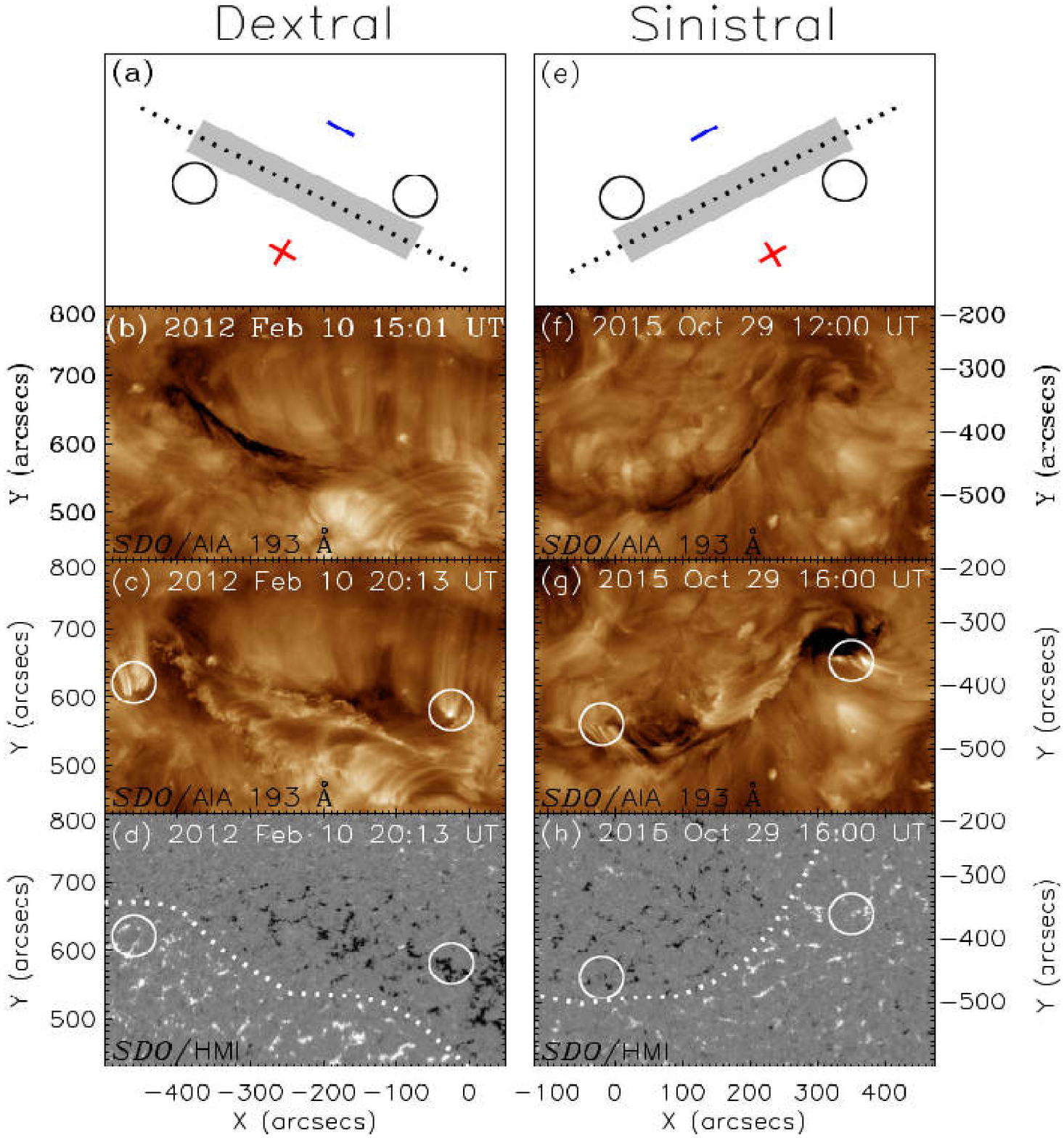}
\caption{Description of our indirect method about how to determine the filament
	chirality by use of the filament draining sites. Top panels are the
	schematic sketches for a dextral (left) and a sinistral (right) filament,
	respectively. Panels (b--c) display the AIA 193 \AA\ evolution of an
	erupting dextral filament, with two circles marking the conjugate draining
	sites, which are then located on the magnetogram in panel (d). It is seen
	that the two draining sites are left-skewed with respect to the magnetic
	PIL (white dashed line). Panels (e--h) are for the case of a sinistral
	filament, where the draining sites are right-skewed.}
	\label{fig1}
\end{figure*}

When a filament erupts, generally two draining sites are visible in EUV images,
and the filament chirality can be identified by the method proposed by
\citet{Che14}, i.e., the chirality of a filament is dextral when the draining
sites are left-skewed, or sinistral when the draining sites are right-skewed.
The application of this method is illustrated in Figure \ref{fig1} and
explained as follows: The top panels are the sketch maps of the cases with left
skew of the draining sites (panel a, corresponding to dextral chirality) and
right skew of the draining sites (panel e, corresponding to sinistral
chirality), respectively, where the dashed lines mark the magnetic PIL, the
shaded areas represent the filament spine, and the two circles in each panel
mark the EUV brightenings associated with the filament drainage. Panels (b--c)
display the evolution of an erupting filament with dextral chirality, whereas
panels (f--g) display the evolution of an erupting filament with sinistral
chirality. To apply this method, we first check the {\it SDO}/AIA images to
obtain the locations of the brightenings associated with the filament draining,
as marked by the circles in panels (c) and (g) of Figure \ref{fig1}. In many
cases, the skew of the draining sites can then be determined. In some cases
where the filament is curved and the two draining sites are too close to the
magnetic PIL, we mark the locations of the draining sites on the {\it SDO}/HMI
magnetogram, as shown by Figure \ref{fig1}(d, h). Here the magnetogram is
derotated to the time of the EUV images. With the correspondence of the
draining sites and the magnetic polarities, we can easily determine the skew
of the draining sites, and hence the chirality of the filament. It is noted
that our chirality identification method works well even when only one
draining site is visible \citep{bi14}.

The validity of this chirality identification method is confirmed with  
small samples by \citet{Ouy15} and \citet{hao16} who found that the results 
are in accordance with the vector magnetograms, the skew of the twin dimmings
and the skew of flaring loops after filament eruption. For a large sample like
the one in this paper, we also checked the chilarity determined by the skew of 
the associated flaring loops in all the disk events, which was proposed by 
\citet{mart95}. It is found that the results obtained by the two methods are 
exactly the same. It is also noted that the bright draining sites of erupting
filaments would not be confused with flaring patches since the brightening is 
preceded by the dark draining filament materials, as indicated by Figures 3 
and 7 in \citet{Ouy15}. For all the events in our sample, we trace the draining 
filament materials to determine the bright draining sites.

\section{Results}\label{sec3}

\begin{figure*}
\epsscale{2.1}
\plotone{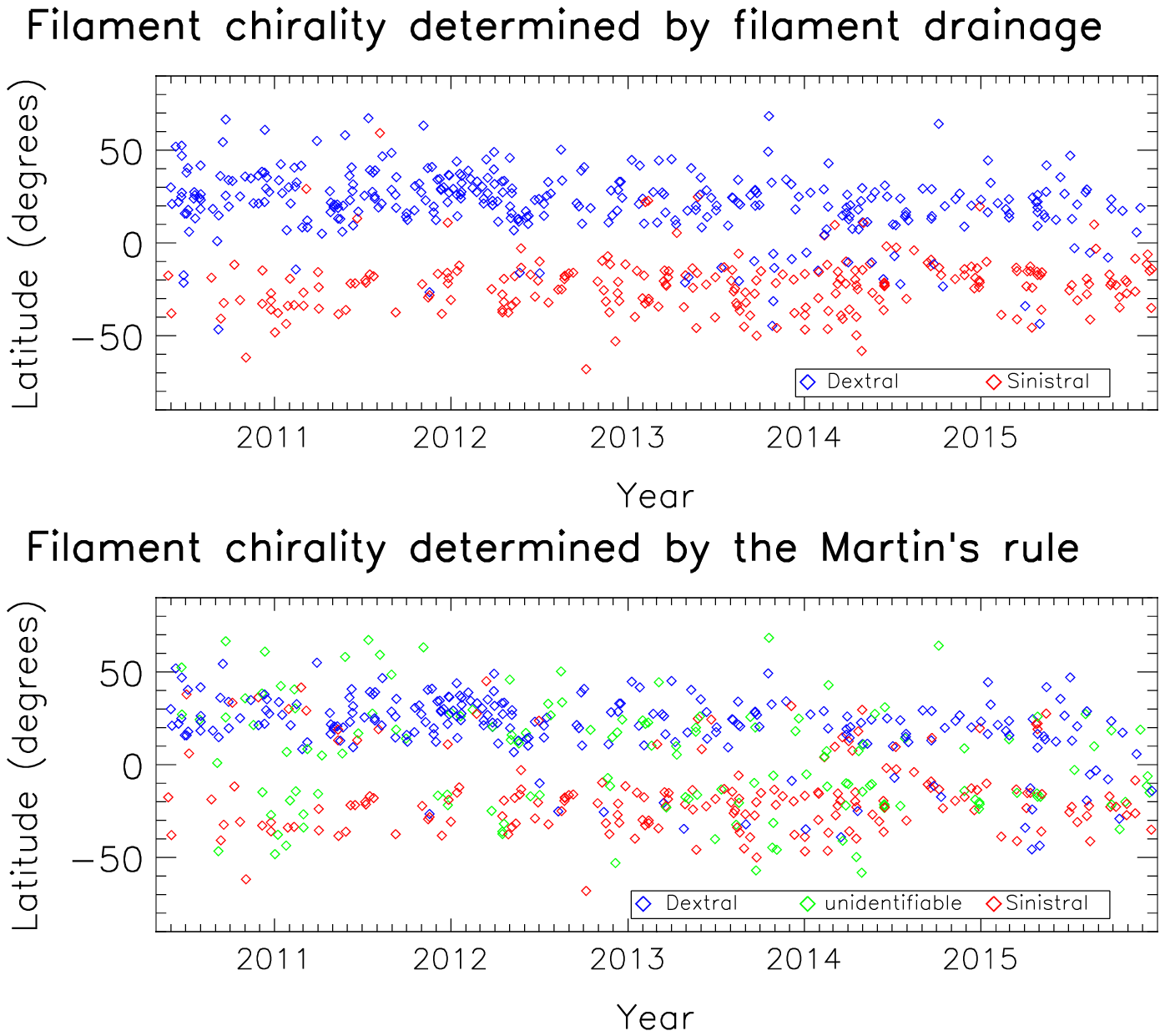}
\caption{Top panel: The time-latitude distribution of chirality for the 571 erupting
	filaments observed from 2010 May to 2015 December, where the chirality is
	determined with the filament drainage method proposed by \citet{Che14},
	and the blue diamonds correspond to dextral chirality, whereas the red
	diamonds to sinistral chirality. Bottom panel: The same as the top panel
	except that the chirality is determined by the bearing sense of filament
	barbs as proposed by \citet{Mar92}, where the filaments without discernable
	H$\alpha$ barbs (hence the filament chirality cannot be determined) are
	labeled with green diamonds.}
  \label{fig2}
\end{figure*}

With the method described in \S\ref{sec2}, the chirality of 571 erupting
filaments is then determined. The top panel of Figure \ref{fig2} plots the
chirality distribution of these filaments, where the horizontal axis is the
time and the vertical axis is the latitude. In this panel, the blue diamonds
correspond to the filaments with dextral chirality (hence negative helicity),
and the red diamonds represent the filaments with sinistral chirality (hence
positive helicity). It is found that 307 out of 324, i.e., 94.8\%, filaments
in the northern hemisphere have negative helicity, and 216 out of 247, i.e.,
87.4\%, filaments in the southern hemisphere have positive helicity. Put
together, 91.6\% of our sample of erupting filaments follow the hemispheric
rule of helicity sign.

In comparison, we apply Martin's rule to the same sample, i.e., to examine the
sign of helicity based on the bearing sense of the filament barbs, which are
observed by {\it GONG} H$\alpha$ telescopes. Among the 571 filaments, 7 events
are not visible in H$\alpha$. Following \citet{Pev03b} and \citet{Jin04}, for
the remaining 564 filaments, the sign of helicity is assigned to be
negative/positive when the filament barbs are predominantly
right-bearing/left-bearing. The resulting distribution of the helicity sign
is displayed in the bottom panel of Figure \ref{fig2} with the same
coordinates as the top panel. Similar to the top panel, the blue diamonds
correspond to the filaments with dextral chirality (hence negative helicity),
and the red diamonds represent the filaments with sinistral chirality (hence
positive helicity). Among all the 564 H$\alpha$ filaments, it is found that
211 out of 322, i.e., 65.5\%, of the filaments in the northern hemisphere have
negative helicity, and 152 out of 242, i.e., 62.8\%, filaments in the southern
hemisphere have positive helicity. It should be noted that 76 filaments in the
northern hemisphere and 66 filaments in the southern hemisphere have no
discernable barbs. If we exclude all the H$\alpha$ filaments without
identifiable barbs, it is found that 85.8\% of the filaments in the northern
hemisphere have negative helicity, and 86.4\% of the filaments in the southern
hemisphere have positive helicity. As a whole, 86.0\% of all the filaments
with clear barbs follow the hemispheric rule of helicity sign, which is 5.6\%
less than the value obtained by our method using the filament draining sites.

Following \citet{Eng98}, we divide the filaments in our sample into three
types, i.e., (1) Quiescent ones, which are located in quiet regions with
relatively weaker magnetic field, (2) Intermediate ones with one end in an
active region and the other in the quiet region, and (3) Active-region ones,
which are located inside an active region. It is found that among the 571
filaments, there are 379 quiescent filaments, 100 intermediate filaments, and
92 active-region filaments. Their time-latitude diagrams of the helicity sign
are displayed in three rows of Figure \ref{fig3}, respectively, where the blue
diamonds correspond to negative helicity, whereas the red diamonds to positive
helicity. It is shown that the strength of the hemispheric rule is 93\% for the
quiescent filaments, 95\% for the intermediate filaments, and 83\% for the
active-region filaments, respectively. On the right side of each row, we sum up
the corresponding type of filaments with time and plot their latitude
distribution in histograms. It is revealed that each type of the filaments have
a bimodal distribution in latitude. Whereas the quiescent filaments are
distributed more broadly in latitude, the other two types are more concentrated
in low latitudes. It is also seen that the quiescent and the
intermediate filaments that are against the hemispheric rule are concentrated 
near the equator, whereas the active-region filaments that are against the 
hemispheric rule are concentrated near the latitude of $\sim$20$^\circ$ in 
each hemisphere.

\section{Discussions}\label{sec4}

\subsection{Strength of the Hemispheric Rule}

It has been argued that the solar dynamo processes generate well-organized
symmetric or antisymmetric magnetic patterns over the two hemispheres of the
Sun \citep[e.g.,][]{zirk97, pevt14}. Important parameters characterizing the
magnetic topology are the helicity and handedness or chirality, which can be
quantified from vector magnetograms and imaging observations. As a result, a
hemispheric rule was revealed, i.e., the helicity tends to be negative in the
northern hemisphere and positive in the southern hemisphere. However, the
strength of the hemispheric rule changes with different proxies and even with
different samples. As summarized by \citet{Wan13}, the strength lies in the
range $\sim$62\%--82\% for sunspot whorls \citep{Hal25, Ric41, Pev03a},
$\sim$60\%--82\% for vector magnetograms of active regions \citep{Pev95,
Abr97, Bao98, Log98, Pev01, Hag05, Zha06}, $\sim$64\%--87\% for X-ray sigmoids
\citep{Rus96, Can99, Lim09}, $\sim$55\%--76\% for active-region filaments
\citep{Mar94, Pev03b}, and $\sim$82\%--84\% for quiescent and intermediate
filaments \citep{Mar94, Pev03b, Lim09}.

It should be pointed out that, for all these proxies, the sign of helicity or
chirality may be mis-identified for different reasons, e.g., projection
effects, Faraday rotation, and so on \citep{xu09}. In terms of the filament
chirality, the traditional identification method is based on Martin's rule,
i.e., right-bearing barbs correspond to dextral chirality, and left-bearing
barbs correspond to sinistral chirality \citep{Mar92}. However, as pointed out
by \citet{Che14}, the Martin's rule is valid only for the inverse-polarity
filaments, i.e., those magnetically supported by a flux rope. For
normal-polarity filaments, i.e., those supported by a sheared arcade, however,
the one-to-one correspondence is opposite, i.e., right-bearing barbs correspond
to sinistral chirality, whereas left-bearing barbs correspond to dextral
chirality. Therefore, when the Martin's rule is applied to a normal-polarity
filament, its chirality would be mis-identified. In order to estimate the
strength of the hemispheric rule of filament chirality, in this paper we
applied a new method proposed by \citet{Che14}, which is based on the skew of
the filament draining sites. It is found that 523 out of 571, i.e., 91.6\%,
erupting filaments during 2010--2015 follow the hemispheric rule. We further
divided these filaments into quiescent, intermediate, and active-region types,
and found that the strength of the hemispheric rule is 93\% for the quiescent
filaments, 95\% for the intermediate filaments, and 83\% for the
active-region filaments, respectively. Compared to the previous studies
\citep{Mar94, Pev03b, Ber05, Yea07, Lim09}, our estimates of the strength for
the quiescent and the intermediate filaments are slightly higher, but our
estimate of the strength for the active-region filaments is remarkably higher.

In order to more quantitatively compare our result with those obtained using
Martin's rule, we also applied the Martin's rule to all our filament events.
The bearing sense of the filament barbs is judged by the H$\alpha$ images
observed by the {\it GONG} network. It is found that there are no H$\alpha$
observations for 7 events. For the remaining 564 filaments, 363, i.e., 64.3\%,
filaments follow the hemispheric rule. It should be noted that among the 564
filaments, 142, i.e., 25.2\%, of them have no clear barbs, hence their
chirality cannot be determined by the Martin's rule. If we exclude these
filaments, then 86.0\% of all the filaments with clear barbs follow the
hemispheric rule. Such a value is about 5.6\% smaller than our estimate using
filament draining sites.

Compared to some of the previous studies, our statistics shows a significantly
stronger hemispheric preference of filament chirality, i.e., dextral in
the northern hemisphere and sinistral in the southern hemisphere. For the
quiescent filaments, \citet{Mar94} analyzed 73 filaments and found that the
strength of the hemispheric rule is 82\%. Later, \citet{Pev03b} analyzed 1436
filaments, which leads to similar strength, i.e., 83\%. However, our estimate
is as high as 93\%. For the intermediate filaments, \citet{Lim09} found that
the strength of the hemispheric rule is 84\%. However, our estimate is up to
95\%. For the active-region filaments, whereas \citet{Pev03b} estimated the
strength of the hemispheric rule to be 76\%, \citet{Mar94} claimed that there
is no hemispheric preference. However, we found that 83\% of the active-region
filaments follow the hemispheric rule. The reason for the remarkable difference
is probably that the previous authors used the bearing sense of filament barbs
to determine the chirality of the filaments, which would lead to
misidentification of chirality when the filament is of the normal-polarity
type, i.e., the corresponding magnetic configuration is a sheared arcade. That
is to say, we cannot judge the chirality of a filament by the bearing sense
of its barbs \citep{Che14, Ouy15}, nor can we judge the chirality by the
magnetic polarities of the two endpoints of a filament \citep{hao16}. On the
contrary, the chirality can be visually determined by other patterns, such as
the skew of the filament draining sites \citep{Che14}, the skew of coronal
loops or flaring loops \citep{mart98}, and the skew of the twin dimmings upon
filament eruptions \citep{jian11}.

\subsection{Cyclic Behavior of the Hemispheric Rule}

It was proposed that the hemispheric rule might be time-dependent
\citep[e.g.,][]{Sak03, Cho04, Hao11, yang12, Gos13}. However, the results were
divergent. In terms of the current helicity, the violation of the hemispheric
rule was claimed to happen in any phase of a solar cycle. For example,
\citet{Sak03}, \citet{Cho04}, and \citet{Hag05} claimed that the hemispheric
rule might be opposite near solar minimum. However, \citet{Bao00} found that
the hemispheric rule is opposite during the rising phase of solar cycle 23.
On the contrary, \citet{Hao11, Hao12, Hao13} suggested that the violation of
the hemispheric rule happens in the declining phase of solar cycle 23.

In terms of the filament chirality, \citet{mack01} theoretically predicted that
the hemispheric rule may disappear during the declining phase of a solar cycle.
\citet{mart14} applied an automated method to a big sample of filaments from
2001 to 2012. As a preliminary result, they showed that the hemispheric rule
waxes and wanes. It is strongly present in 2001--2002 around solar maximum, but
reverses in 2006--2007, which was approaching solar minimum. At other times,
it is wholly absent. Again, they determined the filament chirality by the
Martin's rule, which is valid only for the filaments that are magnetically
supported by a flux rope according to \citet{Che14}.

\begin{figure*}
\epsscale{2.1}
\plotone{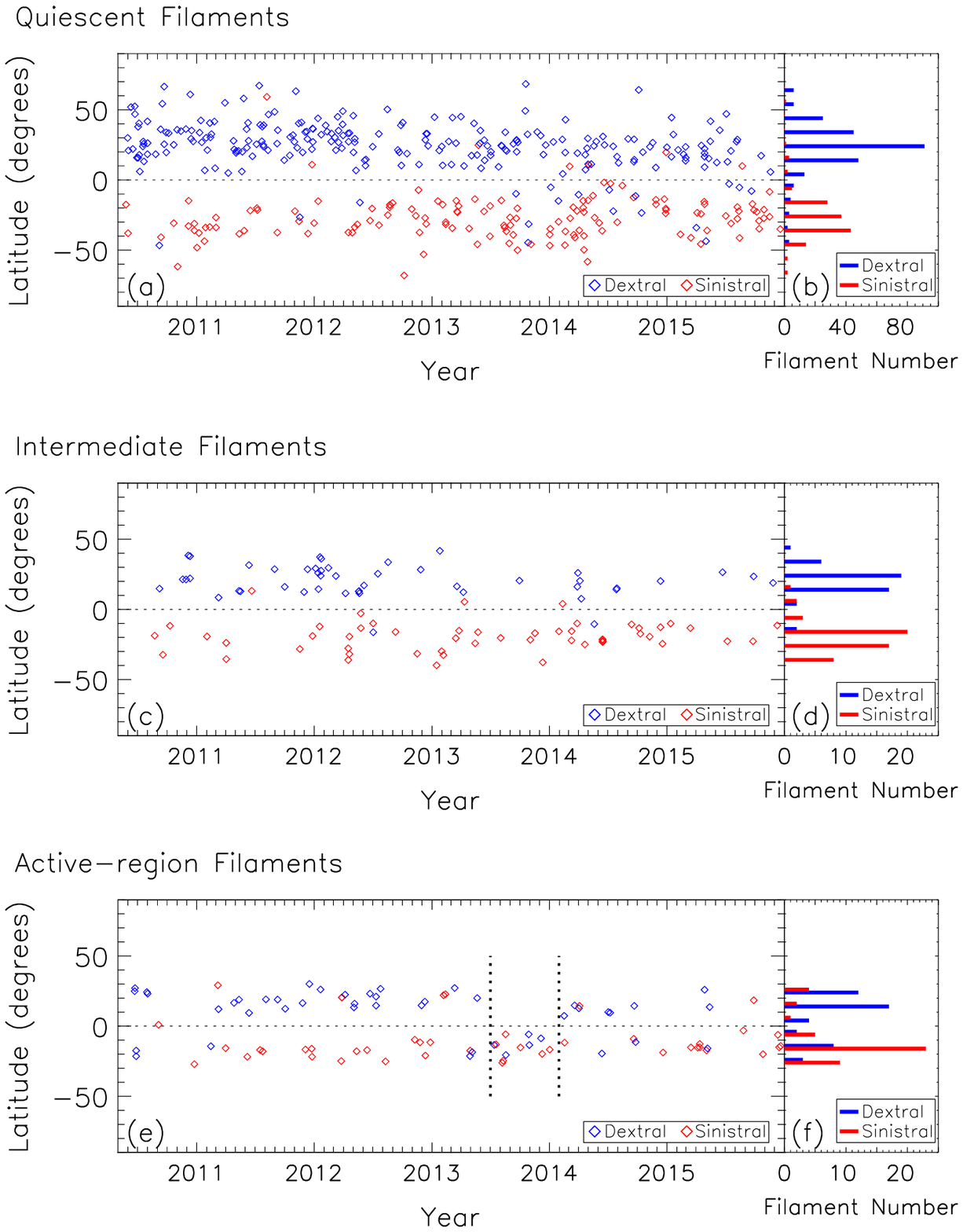}
\caption{The time-latitude distribution of chirality for the 379 quiescent
        filaments (panel a), 100 intermediate filaments (panel c) and 92
	active-region filaments (panel e), where the chirality is determined with
	the filament drainage method, and the blue
	diamonds correspond to dextral chirality, whereas the red diamonds to
	sinistral chirality. On the right side of each row, the filaments are
	summed over time to get the latitude distributions of the corresponding
	dextral and sinistral filaments.}
  \label{fig3}
\end{figure*}

\begin{figure*}
\epsscale{2.1}
\plotone{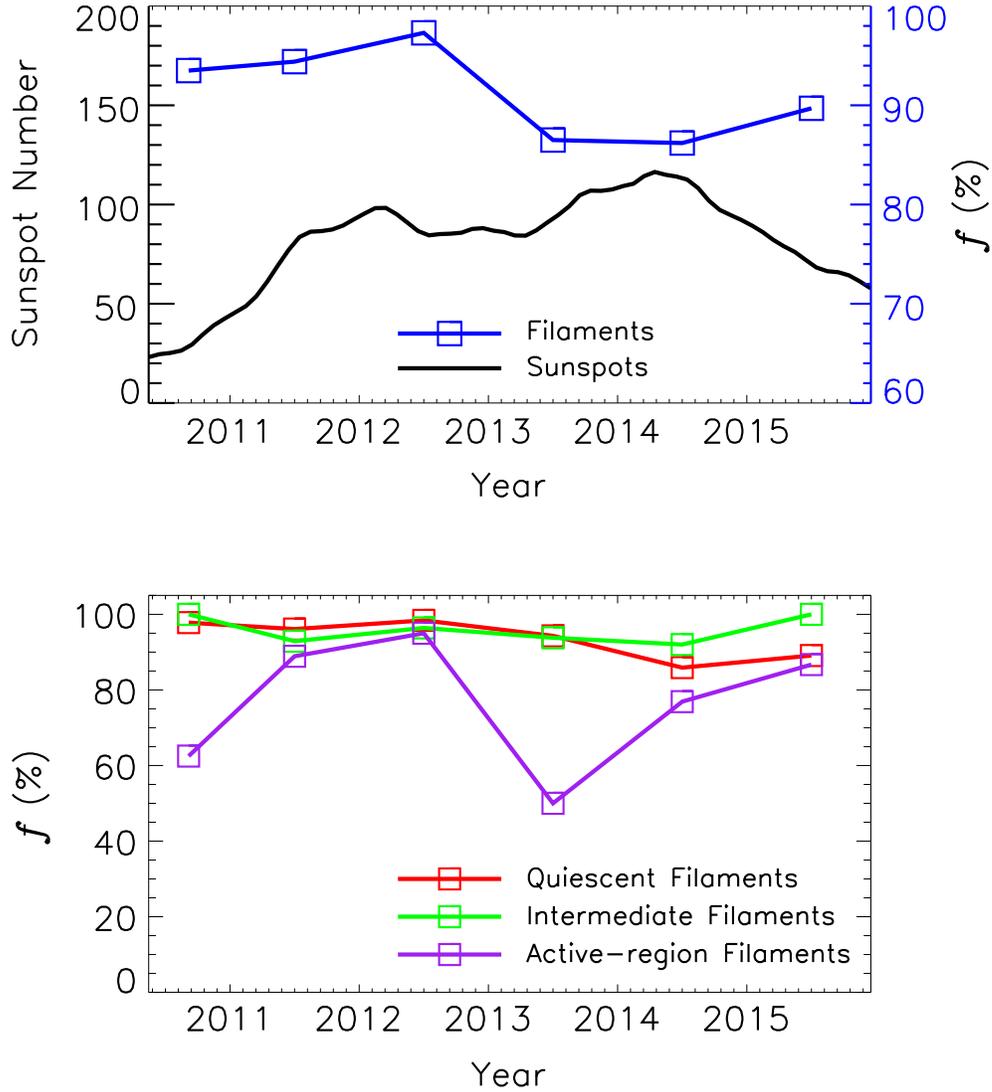}
\caption{Top panel: Cyclic evolution of the sunspot number (black line) and
	$f$, the strength of the hemispheric	rule obeyed by the filament chirality
	(connected blue squares). Bottom panel: Cyclic evolution of $f$ for the
	quiescent filaments (connected red squares), intermediate filaments
	(connected green squares), and active-region filaments ((connected purple
	squares), where a level at 50\% means that the hemispheric preference vanishes.}
  \label{fig4}
\end{figure*}

In this paper, we used the filament draining sites to determine
the filament chirality. A quick look at our results in Figure \ref{fig3}
immediately gives an impression that the hemispheric rule of the filament
chirality roughly holds well from 2010 to 2015. In order to check the cyclic
evolution of the strength of the hemispheric rule, we calculate $f$, the
percentage of the filaments which follow the hemispheric rule each year, and
plot its evolution in Figure \ref{fig4}, where the top panel compares the
strength of the hemispheric rule of all the filaments ({\it connected red
squares}) with the smoothed sunspot number \citep{clet16}, and the bottom panel
compares the strength of the hemispheric rule among quiescent filaments
({\it red squares}), intermediate filaments ({\it green squares}), and
active-region filaments ({\it blue squares}). It can be seen that as a whole,
the filament chirality follows the hemispheric rule very well, and the strength
is $\sim$90\% in the rising phase of solar cycle 24, though slightly decreases
to $\sim$87\% during the solar maximum and the declining phase. After dividing
these filaments into three types, it is then found that the strength of the
quiescent filaments decreases slightly from $\sim$97\% in the rising phase to
$\sim$85\% in the declining phase, whereas the strength of the intermediate
filaments keeps a high value around $96\pm4\%$ from 2010 to 2015. Only the
active-region filaments show significant variations. Their strength of the
hemispheric rule rises from $\sim$63\% to $\sim$95\% in the rising phase, and
keeps a high value $82\pm5\%$ during the declining phase. However, during a
period from 2013 June to 2014 January, which is around the solar maximum, the
hemispheric preference totally vanishes. As seen from Figure \ref{fig3},
whereas no active-region filaments erupt in the northern hemisphere during this
period (marked by the two vertical dashed lines in the bottom panel of Figure
\ref{fig3}), there are equal active-region filaments with dextral and sinistral
chirality in the southern hemisphere. During this period, the sunspot area in
the southern hemisphere reaches its highest peak in solar cycle 24, whereas the
sunspot area in the northern hemisphere reaches a local minimum \citep{Den16}.

Similar to our result, \citet{Bao01} also found that the fraction of active
regions with reversed helicity sign is higher near the solar maximum of solar
cycle 22. We are still not sure whether the absence of the hemispheric rule
of the active-region filaments \citep[and hence the corresponding active
regions,][]{wang13} near solar maximum happens every solar cycle. However, it is
interesting to notice that active regions that do not follow the hemispheric
rule are generally more productive in solar flares \citep{Bao01}.

\subsection{Inverse- and Normal-polarity Filaments}

Magnetic measurements indicate that solar filaments can be divided into
(1) inverse-polarity filaments, whose magnetic field component perpendicular
to the magnetic PIL is opposite to what expected from the photospheric
magnetograms, and (2) normal-polarity filaments, whose magnetic orientation
is the same as expected from the photospheric magnetograms \citep{Ler84,
Bom98}. Theoretically, the two types of filaments are described by the
KR model \citep{Kup74} and the KS model \citep{Kip57}, respectively. The
former corresponds to a flux rope, whereas the latter corresponds to a
sheared arcade. The original 2-dimensional models were later extended to
3-dimensions \citep{Ant94, Aul98, van04}. With solar filaments being the
progenitor of coronal mass ejections \citep[CMEs,][]{Che11}, it is
natural to think that the pre-eruption magnetic structure is a flux rope
in some CME events, and a sheared arcade in other events \citep{gosl99,
Che11, cheng14, song14, cheng15}, which was confirmed by \citet{Ouy15} in
observations. However, we are still lacking the information on the percentage
of the two types of solar filaments.

In 1980s, efforts were made to measure the magnetic field of solar
prominences \citep{Ler84, Bom98}. According to \citet{Ler84}, $\sim$25\%
of their $>$900 measurements in a sample of 120 prominences correspond to
the normal-polarity configuration. However, as they discussed in their
paper, such a ratio is biased by several selection effects. Besides, there
are different numbers of measurements for different prominences. With their
result, we still do not know how many prominences have the normal-polarity
configuration. Unfortunately, there were no systematic measurements of
magnetic field for filaments/prominences in order to distinguish between
inverse-polarity and normal-polarity configurations. However, recently
\citet{Che14} proposed an indirect method to distinguish the two types of
magnetic configurations, i.e., a dextral/sinistral filament with
right-/left-bearing barbs respectively has an inverse-polarity configuration,
whereas a dextral/sinistral filament with left-/right-bearing barbs
respectively has a normal-polarity configuration.

\begin{figure*}
\epsscale{2.1}
\plotone{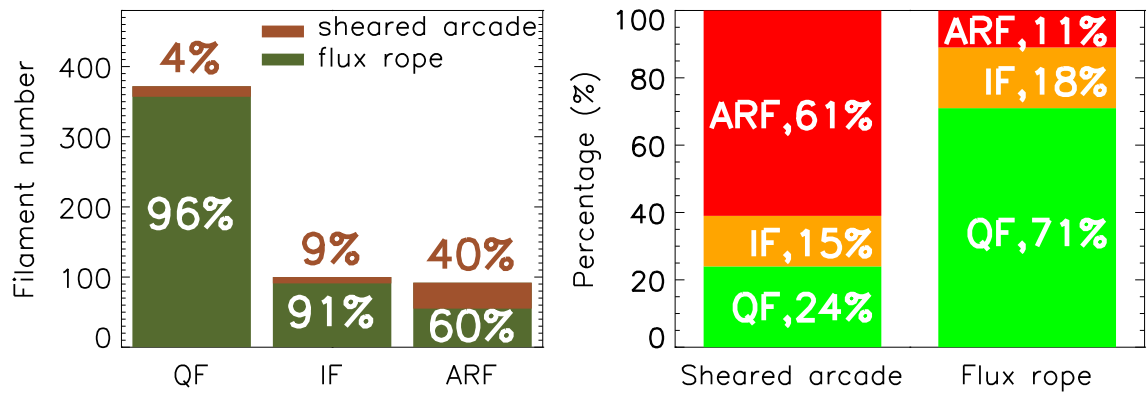}
\caption{Left panel: The fractions of each type of filaments with a sheared arcade
	magnetic configuration (brown) and a flux rope configuration (green), where
	QF stands for quiescent filaments, IF for intermediate filaments, and ARF
	for active-region filaments. Right panel: The fractions of each type of
	magnetic configurations existing in quiescent filaments (green),
	intermediate filaments (orange), and active-region filaments (red).}
  \label{fig5}
\end{figure*}

This method can be summarized as follows: the filaments which obey Martin's
rule have the inverse-polarity configuration, whereas the filaments which
disobey Martin's rule have the normal-polarity configuration. In order to
check the percentage of each type of filaments, we apply this method to our
sample of 571 erupting filaments. Considering that 142 filaments have no clear
barbs, we scrutinize the high-resolution {\it SDO}/AIA images to examine the
bearing sense of the filament threads rather than the filament barbs, since
filament threads are more reliable than filament barbs in determining the
bearing direction \citep{Mar08}. With the filament threads not identifiable in
7 filaments, we finally get a sample of 564 filaments, including 372 quiescent
filaments, 100 intermediate filaments, and 92 active-region filaments. 
It is found that Martin's rule and our new chirality identification 
method agree with each other for 503 out of 564, i.e., 89\%, filaments, i.e., 
these filaments are magnetically supported by a flux rope, therefore are 
inverse-polarity filaments; However, Martin's rule and our new method 
do not agree with each other for 61 out of 564, i.e., 11\%, filaments, i.e., 
they are magnetically supported by a sheared arcade, therefore are
normal-polarity filaments. Among the 61 normal-polarity filaments, there are 15
quiescent filaments, 9 intermediate filaments, and 37 active-region filaments.
In another word, among our sample, 37 out of 92, i.e., 40\%, active-region
filaments are of the normal-polarity type, 9 out of 100, i.e., 9\%,
intermediate filaments are of the normal-polarity type, and 15 out of 372,
i.e., 4\%, quiescent filaments are of the normal-polarity type. These results
are illustrated by the diagrams in Figure \ref{fig5}.

It is noted here that in H$\alpha$ images, many filaments may have
co-existing left-bearing and right-bearing barbs. Some are real, as discussed 
by \citet{guo10}, others might be due to projection effects. In this section, 
we used filament threads to check the bearing sense, instead. It is found that 
only 6 filaments have co-existing left-bearing and right-bearing barbs. We 
take the dominant sense for each filament. It is interesting to notice that 
the minority threads in each filament are generally located near one end of 
the spine, and remain intact upon eruption.

\section{Summary} \label{sec5}

In this paper, we performed statistical analyses on the chirality and the
magnetic configurations (inverse-polarity versus normal polarity) of the solar
filaments which erupt on disk from 2010 May 13 to 2015 December 31, covering
both the rising phase and the beginning of the declining phases of 
solar cycle 24. The chirality is determined by an indirect method proposed by 
\citet{Che14}, i.e.,
left-/right-skewed drainage corresponds to the dextral/sinistral chirality,
respectively. The determination of the magnetic configuration is also based on
a method proposed by \citet{Che14}, i.e., those filaments that follow the
Martin's rule \citep{Mar94} are of the inverse-polarity type, and those that
disobey Martin's rule are of the normal-polarity type. By studying a sample of
571 filaments, we obtained the following results:

(1) About 94.8\% of the filaments in the northern hemisphere have negative
helicity, and 87.4\% of the filaments in the southern hemisphere have positive
helicity, indicating a significant hemispheric preference of helicity. As a
whole, 91.6\% of our sample of erupting filaments follow the hemispheric rule
of helicity sign. With the improved method in determining the filament
chirality, the strength of the hemispheric rule is higher than previous
studies. It should be noted that the statistical result is based on the 
erupting filaments. Those filaments which do not erupt during the disk passage 
are not included in our sample.

(2) Following the conventional way, we divided the filaments into three types,
i.e., quiescent type, intermediate type, and active-region type. It is shown
that the strength of the hemispheric rule is 93\% for the quiescent filaments,
95\% for the intermediate filaments, and 83\% for the active-region filaments,
respectively.

(3) Regarding the cyclic behavior of the hemispheric preference, it is found
that the strength of the quiescent filaments decreases slightly from
$\sim$97\% in the rising phase to $\sim$85\% in the declining phase, whereas
the strength of the intermediate filaments keeps a high value around
$96\pm4\%$ all the time. Only the active-region filaments show significant
variations. Their strength of the hemispheric rule rises from $\sim$63\% to
$\sim$95\% in the rising phase, and keeps a high value $82\pm5\%$ during the
declining phase. However, during a half-year period around the solar maximum,
the hemispheric preference totally vanishes.

(4) It is found that in our sample of erupting filaments, 89\% are
inverse-polarity filaments, which are magnetically supported by a flux rope,
whereas 11\% are normal-polarity filaments, which are magnetically supported
by a sheared arcade.

\acknowledgments
The authors thank the referee for constructive suggestions and the {\it SDO}, 
{\it GONG}, and the Heliophysics Events Knowledgebase (HEK) system teams for 
providing the data. This research was supported by the Chinese foundations 
NSFC (11533005 and 11025314) and Jiangsu 333 Project.

\footnotesize{
\bibliography{reference}
}
\clearpage

\clearpage

\clearpage

\clearpage

\clearpage

\end{document}